\documentclass[manuscript]{aastex61}
\usepackage{amsmath}
\usepackage{amsfonts}
\usepackage{amssymb}
\usepackage{textcomp}
\usepackage{gensymb}
\usepackage{float}
\usepackage{graphicx}
\usepackage{calc}
\usepackage[toc,page]{appendix}
\usepackage[section]{placeins}
\usepackage{morefloats}
\usepackage{mathrsfs}
\usepackage{amsfonts}
\usepackage{silence}
\newif\iffigure
\figuretrue
\newcommand{\alfven}{Alfv\'{e}n}
\renewcommand{\bv}{\boldsymbol{v}}
\newcommand{\bB}{\boldsymbol{B}}
\newcommand{\del}{\partial}
\newcommand{\<}{\langle}
\renewcommand{\>}{\rangle}
\newcommand{\bluewaters}{{\tt blue waters}}

\begin{document}

\title{Resolution Dependence of Magnetorotational Turbulence in the Isothermal Stratified Shearing Box}

\author{Benjamin R. Ryan}
\affil{Department of Astronomy, University of Illinois, 1002 West Green Street, Urbana, IL, 61801}

\author{Charles F. Gammie}
\affil{Department of Astronomy, University of Illinois, 1002 West Green Street, Urbana, IL, 61801}
\affil{Department of Physics, University of Illinois, 1110 West Green Street, Urbana, IL, 61801}

\author{Sebastien Fromang}
\affil{Laboratoire AIM, CEA/DSM-CNRS-Universit\'e Paris 7, Irfu/Service d’Astrophysique, CEA-Saclay, 91191 Gif-sur-Yvette, France}

\author{Pierre Kestener}
\affil{Maison de la Simulation, CEA, CNRS USR3441, Univ. Paris-Sud, UVSQ, Universit\'e Paris-Saclay, 91191 Gif-sur-Yvette, France}

\begin{abstract}

Magnetohydrodynamic (MHD) turbulence driven by the magnetorotational instability can provide diffusive transport of angular momentum in astrophysical disks, and a widely studied computational model for this process is the ideal, stratified, isothermal shearing box.  Here we report results of a convergence study of such boxes up to a resolution of $N = 256$ zones per scale height, performed on \bluewaters{} at NCSA with {\tt ramses-gpu}. We find that the time and vertically integrated dimensionless shear stress $\overline{\alpha} \sim N^{-1/3}$, i.e. the shear stress is resolution dependent. We also find that the magnetic field correlation length decreases with resolution, $\lambda \sim N^{-1/2}$.  This variation is strongest at the disk midplane.  We show that our measurements of $\overline{\alpha}$ are  consistent with earlier studies.  We discuss possible reasons for the lack of convergence.

\end{abstract}

\section{Introduction}

Astrophysical disks form in galaxies and around black holes, neutron stars, white dwarfs, main sequence stars, and planets because the angular momentum of the parent plasma is approximately conserved while kinetic energy in noncircular or non-coplanar motion is easily dissipated and radiated away.  Disk  evolution is therefore governed by angular momentum transport, which can take the form of external torques (e.g. from magnetized winds) or internal transport. 

Diffusive internal transport of angular momentum has been fruitfully described with the phenomenological anomalous viscosity, or $\alpha$, model (\citealt{shakurasunyaev}; \citealt{lyndenbellpringle}), which attributes transport to localized turbulence.  No general driver of turbulence in non-self-gravitating Keplerian disks was known until the discovery by \cite{mri} of the magnetorotational instability (MRI), a local, linear instability of weakly magnetized disks.  Subsequent nonlinear numerical studies convincingly demonstrated that the MRI leads to turbulence and outward angular momentum transport \citep[see the review of][]{balbushawley1998}.  Later work has uncovered purely hydrodynamical instabilities including the zombie vortex (\citealt{zombievortex}, but see  \citealt{lesurlatter}), the vertical shear instability (\citealt{urpin2003}, \citealt{Nelson2013}), the baroclinic instability \citep{Klahr2003, Petersen2007a, Petersen2007b, Lesur2010}, and convective overstability (\citealt{klahr2014}). Nonetheless, MRI-driven turbulence remains the leading candidate for driving disk evolution in many astrophysical settings.

Our paper probes numerical convergence of MHD turbulence in a particular disk model. By convergence, we mean resolution and dissipation-scale independence in average quantities like the angular momentum flux.  We begin by reviewing the various classes of numerical models used to study MHD turbulence in disks, and describing the claims of convergence or nonconvergence made for each class.

Numerical simulations of disk turbulence can be divided into local and global models.  In a {\em local model} (or shearing box; \citealt{goldreich1965}, \citealt{HGB}), the equations of motion are expanded to lowest order in the ratio of the scale height $H$ to the local radius $r$ in a co-orbiting Keplerian frame.  Differential rotation manifests as a linear shear flow.  The shearing box boundary conditions then make it possible to model the disk in a shear-periodic, rectangular box.  The local model is highly symmetric and cannot, for example, distinguish between the inward and outward directions (it is symmetric under a rotation by $\pi$ around the $z$ axis).  In a {\em global model}, by contrast, one simulates some radial range within a disk without requiring $H \ll r$.  Global models do not have the inward-outward symmetry of the local model.

The vertical ($z$) component of gravitational acceleration in the local model is $-\Omega^2 z$, where $\Omega \equiv$ orbital frequency.  {\em Unstratified} local models turn off the vertical component of gravity, begin with a uniform vertical density profile, and typically use periodic vertical boundary conditions.  {\em Stratified} local models turn on the vertical component of gravity, begin with a $z$-dependent vertical density profile, and use a variety of vertical boundary conditions.

For most boundary conditions, local simulations conserve one or more components of the mean magnetic field.  For example, unstratified local models with periodic vertical boundary conditions conserve the mean vertical and toroidal field if the mean radial field vanishes.\footnote{A nonvanishing mean radial field is conserved, but it causes the toroidal field to vary linearly in time.  See \cite{HGB}.}  Numerical investigations show that the mean field can have a profound effect on the saturated turbulent state, so we need to distinguish between {\em zero mean field} models, where all the currents that sustain the field are contained within the simulation volume and can therefore decay, and {\em mean field} models, where one or more components of the field is fixed by the boundary conditions and cannot decay.

Turbulence leads to dissipation.  In {\em explicit dissipation} models (or {\em direct numerical simulations}) dissipation is incorporated directly in the model, for example by a scalar viscosity $\nu$ and resistivity $\eta$ that are dimensionlessly parameterized by the Reynolds numbers and their ratio, the magnetic Prandtl number:
\begin{equation}
Re \equiv \frac{c_s H}{\nu} \qquad
Re_M \equiv \frac {c_s H}{\eta} \qquad
Pr_M \equiv \frac{\nu}{\eta}.
\end{equation}
In {\em implicit large eddy simulation} (ILES) models there is no explicit dissipation, and dissipation is provided by the numerical scheme through truncation error at the grid scale.  Notice that for ILES models run with a conservative scheme, lost kinetic and magnetic energy is entirely captured as plasma thermal energy.  In this sense reconnection can be ``included'' in an ILES model, although the reconnection rate and dynamics may be incorrect.

The consequences of using ILES to study high Reynolds number hydrodynamic turbulence are fairly well understood (e.g.~\citealt{sagaut2006}): if there is sufficient dynamic range (large enough zone number) then the character of dissipation at small scales has little influence on turbulent structures at large scales.  It is large-scale structures that often determine the flow properties of greatest astrophysical interest, such as turbulent momentum flux.  The consequences of using ILES to study high Reynolds numbers magnetohydrodynamic (MHD) turbulence are less well understood (\citealt{miesch2015}).  It is fair to say that many disk simulators (including us) have frequently {\em assumed} that with enough dynamic range MHD ILES would converge to the astrophysically relevant high Reynolds numbers result (but see \citealt{lesur2007}, \citealt{fromangpap2007}, \citealt{longaretti2010}, \citealt{meheut2015}, \citealt{walker2016}).

Finally, disk simulations can be subdivided according to their treatment of heating and cooling of the plasma. Direct simulation of the interaction of radiation with matter has, until recently, been expensive in comparison to available computational resources.  Most disk simulations have therefore used simplified treatments of plasma thermodynamics, with phenomenological cooling and heating, or assumed an isothermal equation of state with pressure $P = \rho c_s^2$, and $c_s$ constant in time and space.  Isothermal models are relevant to disks heated by external illumination, such as disks around compact objects at many gravitational radii, where the thermal timescale can be short compared to the dynamical timescale.

Local models also depend on the box dimensions which are purely numerical parameters. Changes in box sizes are known to produce qualitative changes in shearing box models (e.g.\ \citealt{simon2012}, \citealt{shi2016}). Even the largest domains find correlations on the scale of the box, at least in the corona (\citealt{guan2011}). Two related questions emerge. Does the shearing box model converge as the box sizes goes to infinity? Does shearing box  evolution match global behavior as the box size goes to infinity? These questions are challenging to answer numerically.

Much is now understood about convergence of the gross, time-averaged properties of MRI-driven turbulence (e.g. $\alpha$) in every corner of the five dimensional disk model parameter space: local/global, stratified/unstratified, mean/zero net field, ILES/explicit dissipation, isothermal/nonisothermal. A summary of previous calculations emphasizing convergence is given in Table \ref{tab:prevcalcs}.

\begin{deluxetable}{cccccccc}
\tablewidth{0pc}
%\tablewidth{8in}
% No blank lines inside deluxetable
\rotate % landscape
\tabletypesize{\scriptsize} % 11 pt, \footnotesize for 10pt, nothing for 12pt
\tablehead{
\colhead{Geometry} &
\colhead{Stratified} &
\colhead{Net} &
\colhead{Dissipation} &
\colhead{Isothermal} &
\colhead{Convergent} &
\colhead{Maximum} &
\colhead{References} \\[-0.1in]
\colhead{} &
\colhead{} &
\colhead{Flux} &
\colhead{} &
\colhead{} &
\colhead{} &
\colhead{Resolution} &
\colhead{}
}
\tablecaption{{\footnotesize Convergence Properties of MHD Disk Turbulence Models}\label{tab:prevcalcs}}
\startdata
    local  & no  & zero & ILES     & yes & no        & $256/H$                    & (1) (2) (3) (4) \\[-0.1in]
    local  & no  & zero & explicit & yes & yes       & $512/H$                    & (5) \\[-0.1in]
    local  & no  & mean & ILES     & yes & yes       & $256/H$                    & (6) (2) (3) \\[-0.1in]
    local  & no  & mean & explicit & yes & yes       & $800/H$                    & (7) \\[-0.1in]
    local  & no  & mean & ILES     & no  & yes       & $64/H$                     & (8) \\
    \hline \hline \\[-0.35in]
    local  & yes & zero & ILES     & yes & this work & $256/H$                    & (9) (10) (11) \\[-0.1in]
    local  & yes & zero & ILES     & no  & unclear   & $64/H$                     & (12) (13) \\[-0.1in]
    local  & yes & zero & explicit & yes & unclear       & $128/H$                    & (14) (15) \\[-0.1in]
    local  & yes & mean & ILES     & yes & unclear   & $48/H$                     & (16) (17) \\[-0.1in]
    local  & yes & zero & ILES     & no  & unclear   & $64/H$                     & (18) (19) \\
    \hline \hline \\[-0.35in]
    global & no  & zero & ILES     & yes & yes       & $480\times 1920\times 128$ & (20) \\[-0.1in]
    global & yes & zero & ILES     & no  & unclear   & $768\times 256\times 256$  & (21) (22) (23) \\[-0.1in]
    global & yes & mean & ILES     & no  & unclear   & $288\times 128\times 128$  & (24) (25) (26) \\
\enddata
\tablerefs{{\scriptsize (1) \cite{fromang2007}; (2) \cite{simonhawleybeckwith2009}; (3) \cite{guan2009}; (4) \cite{bodo2011}; (5) \cite{fromang2010}; (6) \cite{HGB}; (7) \cite{meheut2015}; (8) \cite{jiang2013}; (9) \cite{davis2010}; (10) \cite{bodo2014}; (11) \cite{nauman2014}; (12) \cite{shi2010}; (13) \cite{bodo2015} (14) \cite{simon2011}; (15) \cite{oishi2011} (16) \cite{bai2013}; (17) \cite{fromang2013}; (18) \cite{jiang2013inst}; (19) \cite{bodo2015};  (20) \cite{sorathia2012}; (21) \cite{shiokawa2012}; (22) \cite{hawley3D}; (23) \cite{parkin}; (24) \cite{sashamad}; (25) \cite{mckinneymad}; (26) \cite{beckwith2009}}}
\tablecomments{{\scriptsize For convergence (stress with respect to dissipative scale), {\it no} and {\it yes} indicate clear, consistent, persuasive findings in the literature.  This table is incomplete: it focuses on studies that consider convergence, and omits some combinations of parameters. A global unstratified simulation has cylindrical geometry and neglects vertical gravity. $H \equiv c_s / \Omega$.}}
\end{deluxetable}

Zero net field, local, unstratified, isothermal, ILES models are particularly interesting: \cite{fromang2007} showed that these models are nonconvergent (see also \citealt{pessah2007}), and this has been independently confirmed (\citealt{simonhawleybeckwith2009},  \citealt{guan2009}).  With $N$ the number of resolution elements along one axis, with zone aspect ratios fixed, nonconvergence appears as $\alpha \propto N^{-1}$ (but see \citealt{bodo2011}) and magnetic correlation length $\lambda \propto N^{-1}$ (i.e. correlation length is proportional to zone size). But this is not the full story: \cite{shi2016} have recently found convergence if the vertical extent of the model is large compared to the radial extent.  In this case MHD turbulence excites waves that travel vertically, and this may be connected to the butterfly oscillations seen in stratified models. However, the connection between these tall boxes and traditional unstratified (and stratified) shearing boxes is still uncertain, and we consider it premature to change the relevant conclusion for convergence in Table \ref{tab:prevcalcs}.

Unstratified models converge, however, if either explicit dissipation (\citealt{fromang2010}, but see \citealt{bodo2011}) or a mean magnetic field (\citealt{simonhawleybeckwith2009}, \citealt{guan2009}) are added.  When a mean field is added $\alpha$ increases proportional to the mean field strength (\citealt{HGB}, \citealt{salvesen2015}).

What about stratified models?  One might think that stratification would lead to magnetically driven convection which could organize the field on the scale of the convective eddies, leading to convergence.  But the numerical evidence for convergence of zero net field, local, stratified, isothermal ILES models is contradictory.  The work of \cite{davis2010}, using the {\tt athena} code, is consistent with convergence, while the work of \cite{bodo2014}, using the {\tt pluto} code, shows a sharp drop in Maxwell stress at the highest accessible resolution of $200/\sqrt{2} \simeq 141$ zones per scale height.  The question of convergence for stratified, isothermal ILES models is particularly pressing because they are sometimes used to interpret observations in both local (e.g.\ \citealt{simon2015}) and global (e.g.\ \citealt{moscibrodzka2009}, \citealt{flock2015}) forms.

This paper therefore returns to study the convergence of zero net field, local, stratified, isothermal ILES models at high resolutions made newly accessible by the combination of NCSA's \bluewaters{} machine and the {\tt ramses-gpu} code.  In Section 2 we present the physical model and numerical method.  Section 3 contains the results of our calculations. Section 4 discusses the implications of our results and future directions. Section 5 concludes.

\section{Model}

\subsection{Governing Equations}

The local model expands the equations of motion to lowest nontrivial order around a Keplerian orbit at $R = R_0, \phi = \phi_0 + \Omega t, z = 0$ and defines the local Cartesian coordinates
\begin{equation}
\left(x, y, z \right) = \left( R-R_0, R_0 \left(\phi - \Omega t - \phi_0 \right), z\right).
\end{equation}
In the local model for a Keplerian disk the equations of ideal MHD, with an isothermal equation of state ($P = \rho c_s^2$; $P \equiv$ pressure, $\rho \equiv$ density, $c_s \equiv$ sound speed, which is assumed constant), are  
\begin{align}
\frac{\partial \rho}{\partial t} &= -\nabla \cdot \left( \rho \bv\right), \\
\frac{\partial \rho \bv}{\partial t} 
&= -\nabla \cdot \left(\rho \bv\bv \right) + 
\left( \bB \cdot \nabla \right) \bB 
- \nabla \left( \frac{\bB \cdot \bB}{2}+ P\right)\label{momeq}\\
& \qquad\qquad  -2 \, \rho \, \Omega \hat{\boldsymbol{e}}_z \times \bv  +
\rho \nabla \left( -\frac{3}{2}\Omega^2 x^2 + \frac{1}{2}\Omega^2 z^2\right) \notag, \\
\frac{\partial \bB}{\partial t} &= \nabla \times \left( \bv \times \bB\right),
\end{align}
where $\bv \equiv$ velocity in the local frame and $\bB \equiv$ magnetic field, subject to the constraint
\begin{equation}
\nabla \cdot \boldsymbol{B} = 0.
\end{equation}
Equation (\ref{momeq}) includes Coriolis and tidal forces.  Notice that there is no explicit dissipation (resistivity or viscosity) and that $R_0$ does not appear in the governing equations. 

For $\boldsymbol{B} = 0$ these equations admit the equilibrium
\begin{align}
\rho &= \rho_0 \exp \left( -\frac{z^2}{2 H^2}\right),\label{eqn:rhoinit} \\
\boldsymbol{v} &= -\frac{3}{2}x \Omega  \hat{\boldsymbol{e}}_y.
\label{eqn:vinit}
\end{align}
Here $H = c_s/\Omega$.  Notice that others (e.g.~\citealt{davis2010}, \citealt{bodo2014}) define the scale height as $\sqrt{2} c_s/\Omega$.  This implies that their resolution should be multiplied by $1/\sqrt{2}$ for comparison with ours.  The initial conditions for our model are the unmagnetized equilibrium (\ref{eqn:rhoinit})-(\ref{eqn:vinit}), seeded with a uniform toroidal field at $|z| < 2H$; ${\bf B} = 0$ elsewhere.  The initial plasma $\beta \equiv 2 P/B^2 = 50$ at the midplane.

Hereafter we set
\begin{equation}
c_s = 1 \qquad \Omega = 1 \qquad \rho_0 = 1
\end{equation}
which together imply that $H = 1$ and the surface density $\Sigma = \sqrt{2\pi}$.  The mass, length, and time units are thus $\rho_0 H^3$, $H$, and $\Omega^{-1}$, respectively.  Occasionally we reinsert these units for clarity. 

For the $x$ and $y$ boundaries we use {\em shearing box boundary conditions} \citep[see][]{HGB}.  With these boundary conditions the model is translation-invariant in the $x-y$ plane, and also invariant under rotations by $\pi$ around the $z$ axis. In addition, the vertical magnetic flux $\Phi_z \equiv \int dx dy B_z$ (integral taken over the entire $x-y$ domain at any $z$) is conserved.
Our initial conditions have $\Phi_z = 0$, and our model extends over $-L_x/2 < x < L_x/2$, with $L_x = 3$, and over $-L_y/2 < y < L_y/2$, with $L_y = 4$.

At least three different $z$ boundary conditions have been used for stratified shearing boxes.  Beginning with \cite{stone1996}, many have used outflow boundary conditions. \cite{davis2010} used periodic boundary conditions, which have the advantage that all three components of the mean magnetic field are conserved in exchange for altering the domain topology. Several authors (\citealt{brandenburg1995}, \citealt{bodo2014}) adopt impenetrable, stress-free boundaries that set $\partial v_x / \partial z = \partial v_y / \partial z = v_z = 0$ and $B_x = B_y = \partial B_z / \partial z = 0$ (or the equivalent conditions on the magnetic vector potential).  The effect of boundary condition choice has not been systematically studied at modern resolution, although \cite{stone1996} found that an artificial resistive layer at $2 < |z| < 3$ did not affect midplane dynamics significantly, and \cite{oishi2011} demonstrate similar behavior in three runs that differ only by choice of vertical boundary conditions. For finite thermal diffusivity, \cite{gressel2013} find a significant change in energy transport between outflow and impenetrable vertical boundaries.

We chose outflow boundary conditions and a large vertical extent to minimize the influence of the vertical boundaries on the body of the disk. Formally, outflow boundary conditions are $\del {\bf B}/\del z = 0$ and $\del{\bf v}/\del z = 0$, and $-(1/\rho)\del(c_s^2\rho)/\del z - \Omega^2 z = 0$, consistent with hydrostatic equilibrium.  Our model extends over $-L_z/2 < z < L_z/2$ with $L_z = 12$.

Outflow boundary conditions imply that fluid mass in the computational domain is not conserved.  The  characteristic outflow timescale $\tau_{\mathrm{out}} \equiv \Sigma /\dot{\Sigma}$.  Assuming sonic outflow at the boundaries, $\dot{\Sigma} \approx 2 \rho(|z| = L_z/2) c_s$.  Using the density profile fit from \cite{guan2011} that takes account of magnetic support of the disk atmosphere, $\tau_{\mathrm{out}} \approx 6 \times 10^{4}\Omega^{-1}$.  This is long compared to our integration times.  Outflow boundary conditions also imply that the radial and toroidal magnetic flux are not conserved.

The domain size $(L_x,L_y,L_z) = (3, 4, 12) H$ may influence the turbulent state.  \cite{guan2011} and \cite{simon2012} provide numerical evidence that angular momentum transport and variability may depend on structures large compared to $H$, but such large domains are currently inaccessible at our target resolution. \cite{simon2012} demonstrate a transition to anomalous behavior as $L_x$ goes from $2$ to $0.5$. The minimum $L_x$ that avoids these pathologies is known to be less than $L_x = 3$; the results of \cite{davis2010} suggest that this minimum is less than $L_x = \sqrt{2}$.

Finally, the integration time $\Delta t$ should be long enough that average values for $\alpha$ and other quantities of interest can be measured with reasonable signal to noise. Our typical integration time is $\approx 300 \Omega^{-1}$ (see the Appendix for a discussion of measurement errors).

\subsection{Numerical Methods}
\label{sec:numerics}

We integrate the model with {\tt ramses-gpu} \citep{fromang2006, kestener2010, kestener2014}, a modern astrophysical MHD code with support for GPU acceleration\footnote{Freely available: \url{http://www.maisondelasimulation.fr/projects/RAMSES-GPU/html/}}. {\tt ramses-gpu} is a second-order finite volume MUSCL scheme. Fluxes are evaluated with the HLLD approximate Riemann solver (\citealt{hlld}). The constraint $\nabla \cdot {\bf B} = 0$ is preserved via constrained transport with face-centered magnetic fields (\citealt{evanshawley1988}).

Numerical resolution is characterized by
\begin{equation}
N = \frac{H}{\Delta x}
\end{equation}
i.e. the number of zones per scale height in the radial direction.  We take $\Delta x: \Delta y: \Delta z = 1:2:1$, so this is also the number of zones per scale height in the vertical direction, and twice the number of zones per scale height in the azimuthal direction. \cite{hawley2011} showed that for MRI growth the azimuthal direction is typically better resolved than the vertical direction by a factor of a few in shearing boxes, as did \cite{parkin}. \cite{guan2009} showed that the autocorrelation function of the magnetic field in unstratified, isothermal shearing box models is anisotropic and approximately in the ratio $1:4:1$ in the radial, azimuthal, and vertical directions, suggesting that near the midplane the $y$ direction is slightly better resolved than $x$ and $z$ in our model.

The mean azimuthal velocity $\overline{v}_y = -(3/2)\Omega x$.  Truncation error depends on the velocity of the fluid with respect to the grid, and therefore if $\overline{v}_y$ is the dominant component of the velocity field the truncation error will vary systematically with $x$.  This problem can be solved by using orbital advection (also known as ``FARGO''; \citealt{masset2000}) for the MHD equations (\citealt{johnson2008}, and references therein).  We do not use orbital advection, but the shear velocity at the edge of our boxes is only $1.5 c_s$, and we have checked that the Maxwell and Reynolds stresses do not vary significantly with $x$.

We start preliminary models from smooth initial conditions.  These were seeded with white noise, with $\delta v_i \sim 0.01 c_s$ and $\delta \rho \sim 0.01 \rho_0$ to excite a spectrum of unstable modes.  We used late-time snapshots from these models to initialize our production models.  Each run at resolution $2N$ was initialized with a snapshot from the final (or near-final) state of a model with resolution $N$ using a divergence-free prolongation operator \citep{fromang2006}.  While this avoids running high resolution models through an initial transient phase (and allows our model to forget the initial net azimuthal magnetic flux), it does introduce a potential bias by coupling the initial state of one simulation to the final (or near-final) state of a lower resolution model.

Stratified shearing box models have high \alfven{} speeds in the upper atmosphere ($v_A \sim \rho^{-1/2}$), which via the Courant condition can demand an impractically small timestep.  This is a standard problem in numerical MHD, and can be solved by applying a density floor, or re-introducing a displacement current that limits the \alfven{} speed to a maximum speed \citep{Boris1970}.  In shearing box models, \cite{millerstone2000} used a version of the Boris fix with speed of light $v_{A,max} = (1,4,8) c_s$. \cite{guan2011}, by contrast, impose a density floor of $10^{-5}\rho_0$.  We impose a density floor such that $v_A < v_{A,max} = 10 c_s$.  Our $v_{A,max}$ is higher than the expected $v_A$ at $z = 6$ (as deduced from the fit to averaged stratified shearing box properties of \citealt{guan2011}) but small enough to limit the integration to a practical timestep.

\begin{deluxetable}{cccc}
% No blank lines inside deluxetable
%\tabletypesize{\footnotesize} % 11 pt, \footnotesize for 10pt, nothing for 12pt
\tablewidth{0pc}
\tablehead{
\colhead{Label} &
\colhead{N = Zones/$H$} &
\colhead{$t_0 (\Omega^{-1})$} &
\colhead{$\Delta t (\Omega^{-1})$}
}
\tablecaption{Model parameters\label{tab:parms}}
\startdata
    {\tt r32}  & 32  & 1800 & 300 \\
    {\tt r64}  & 64  & 2100 & 300 \\
    {\tt r128} & 128 & 2400 & 300 \\
    {\tt r256} & 256 & 2648 & 288 \\
\enddata
\end{deluxetable}

\begin{deluxetable}{ccccccccc}
% No blank lines inside deluxetable
%\tabletypesize{\footnotesize} % 11 pt, \footnotesize for 10pt, nothing for 12pt
\tablewidth{0pc}
\tablehead{
\colhead{Label} &
\colhead{$\overline{\alpha}$} &
\colhead{$\overline{\langle -B^x B^y \rangle}$} &
\colhead{$\overline{\langle \rho v^x \delta v^y \rangle}$} &
\colhead{$\Delta M / M_0$} &
\colhead{$\sigma_{\alpha} / \overline{\alpha}$} &
\colhead{$\lambda_{\mathrm{minor}}$} &
\colhead{$\lambda_{\mathrm{major}}$} &
\colhead{$\theta_{\mathrm{tilt}}$}
}
\tablecaption{Model results\label{tab:results}}
\startdata
    {\tt r32} & 0.039  & 0.0061 & 0.0017 & 0.69\% & 0.24 & 0.12  & 0.61  & 16.0\degree \\
    {\tt r64} & 0.034  & 0.0053 & 0.0015 & 0.65\% & 0.37 & 0.085 & 0.40  & 17.8\degree \\
    {\tt r128} & 0.025 & 0.0039 & 0.0011 & 0.55\% & 0.26 & 0.060 & 0.27 & 18.6\degree \\
    {\tt r256} & 0.019 & 0.0029 & 0.0008 & 0.40\% & 0.23 & 0.043 & 0.20 & 19.0\degree \\
\enddata
  \tablecomments{$\lambda_{\mathrm{minor}}$, $\lambda_{\mathrm{major}}$, and $\theta_{\mathrm{tilt}}$ are averaged over $|z| < 2H$.}
\end{deluxetable}

In characterizing the saturated state, we use the following averages: an average over volume
\begin{equation}
\langle f \rangle \equiv \frac{\int d x d y d z \, f}{\int d x d y d z},
\label{eqn:volavg}
\end{equation}
an average over $x$ and $y$
\begin{equation}
[ f ] \equiv \frac{\int d x d y \, f}{\int d x d y},
\end{equation}
and an average over time
\begin{equation}
\overline{f} \equiv \frac{\int d t \, f}{\int d t}.
\label{eqn:timeavg}
\end{equation}
The height-integrated Shakura-Sunyaev $\alpha$ parameter is 
\begin{equation}
\alpha \equiv \frac{\langle \rho v^x \delta v^y - B^x B^y \rangle}{\langle P \rangle}.
\label{eqn:alphadef}
\end{equation}
This definition does not depend explicitly on box size.  It is the average used in height-integrated disk evolution models (e.g.\ \citealt{king2007}) for comparison with observation.

\section{Results}
\label{sec:results}

We consider four models, marching forward in linear resolution by factors of two from $N = 32$ to $N = 256$.  Each model is started using late-time data from the preceding lower-resolution model.  All share a common coordinate time $t$. The runs and their linear resolution, initial time $t_0$, and duration $\Delta t$ are given in Table \ref{tab:parms}. We define $t^{\prime}$ for each run as $t-t_0$. Poloidal slices from all four resolutions are shown in Figure \ref{fig:coolcolorimage}.

\iffigure
\begin{figure}      % use "figure*" instead of "figure" if you want your figure to span both columns
%\epsscale{1.5}      % adjust this number to change the size of your figure
\plotone{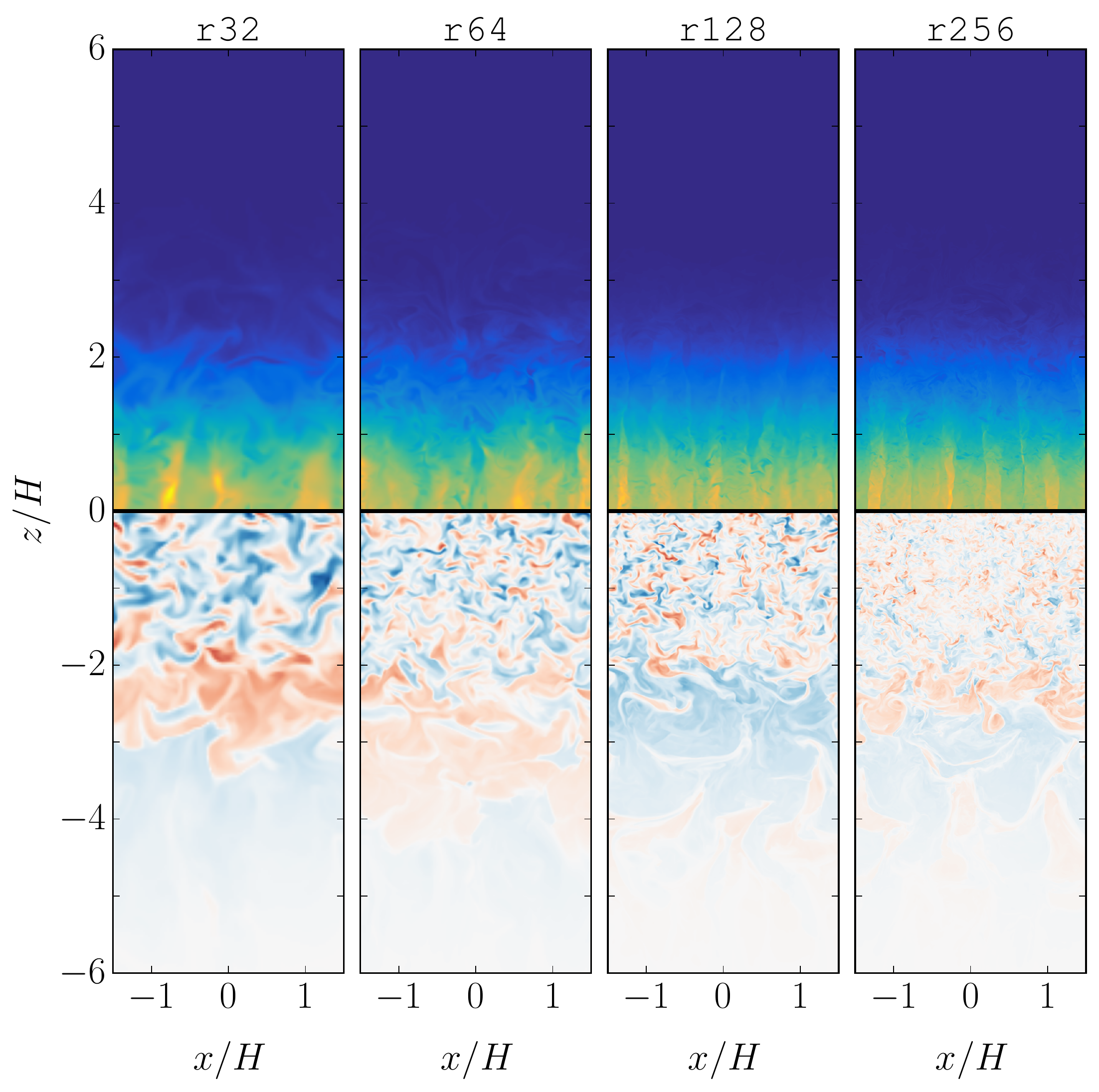}
\centering
\caption{$x$-$z$ slices of $\rho$ (upper half) and $B^y$ (lower half) for 32, 64, 128, and 256 zones per scaleheight. Note that as resolution increases, shocks become sharper and magnetic field structure becomes smaller. Color maps are linear and shared across resolutions.}
\label{fig:coolcolorimage}
\end{figure}

For $L_z = 12 H$ about $0.5\%$ of the disk mass is lost per $300 \Omega^{-1}$ after accounting for mass added via the density floor (see Table \ref{tab:results}; $M_0$ is the mass of the disk at the start of that run).

We now turn to the effects of resolution on one- and two-point statistics of the saturated state. Section \ref{sec:avgstress} considers volume- and area-averaged quantities over the domain. Section \ref{sec:corr} presents the correlation function of the magnetic field.

\subsection{Space and Time Averages}
\label{sec:avgstress}

Does $\overline{\alpha}$ depend on resolution? Figure \ref{fig:alpha} shows $\alpha$ as a function of time and resolution. Average values are given in Table \ref{tab:results}.  Interestingly, the stress monotonically decreases with resolution and there is no evidence for convergence. The resolution dependence is well fit by $\overline{\alpha} \propto N^{-1/3}$.

How large are the error bars on our estimate of $\overline{\alpha}$, and is the observed variation with $N$ significant?  We assume that $\alpha(t)$ is a stationary process with mean $\overline{\alpha}$ and variance $\sigma_\alpha^2$.  We provide evidence in the Appendix that the fluctuations in $\alpha(t)$ decorrelate over large time intervals for a long-integration-time $N=32$ model, and that the correlation time $\tau_c\Omega \approx 63$.  A measurement of $\overline{\alpha}$ averaged over some interval $T$ therefore consists of approximately $\sim T/\tau_c$ independent measurements, and one expects an rms error in evaluating $\overline{\alpha}$ of $\approx \sigma_\alpha (T/\tau_c)^{-1/2}$ \citep[see Fig 4 of][which implies $\tau_c\Omega \sim 10$ in an unstratified local model]{longaretti2010}.  

In the Appendix we work out the relation between $\sigma, \tau_c$ and the rms error in evaluating $\overline{\alpha}$ for a class of model power spectra, assuming $\alpha(t)$ is a Gaussian process.\footnote{The PDF of $\alpha$ {\em is not} consistent with a Gaussian.  The PDF of $\log\alpha$ {\em is} consistent with a Gaussian.  The analysis in the appendix does not change if carried out for $\log\alpha$ instead of $\alpha$}  For a fit to the $N = 32$ run power spectrum, these imply that the expected rms error is $\approx 0.6 \sigma_\alpha/\alpha_0 \approx 0.17$, assuming that $\sigma_\alpha/\alpha_0$ is independent of $N$, consistent with Table \ref{tab:results}.  This can be compared to $\overline{\alpha}(N)/\overline{\alpha}(2N) - 1 \approx 0.25$.  Therefore  the observed trend over a factor of $8$ in $N$ and $\approx 2$ in $\overline{\alpha}$ is significant.  A naive estimate of the probability that $d\log\overline{\alpha}/d\log N \ge 0$ gives $\approx 3\%$.

\begin{figure}      
\plotone{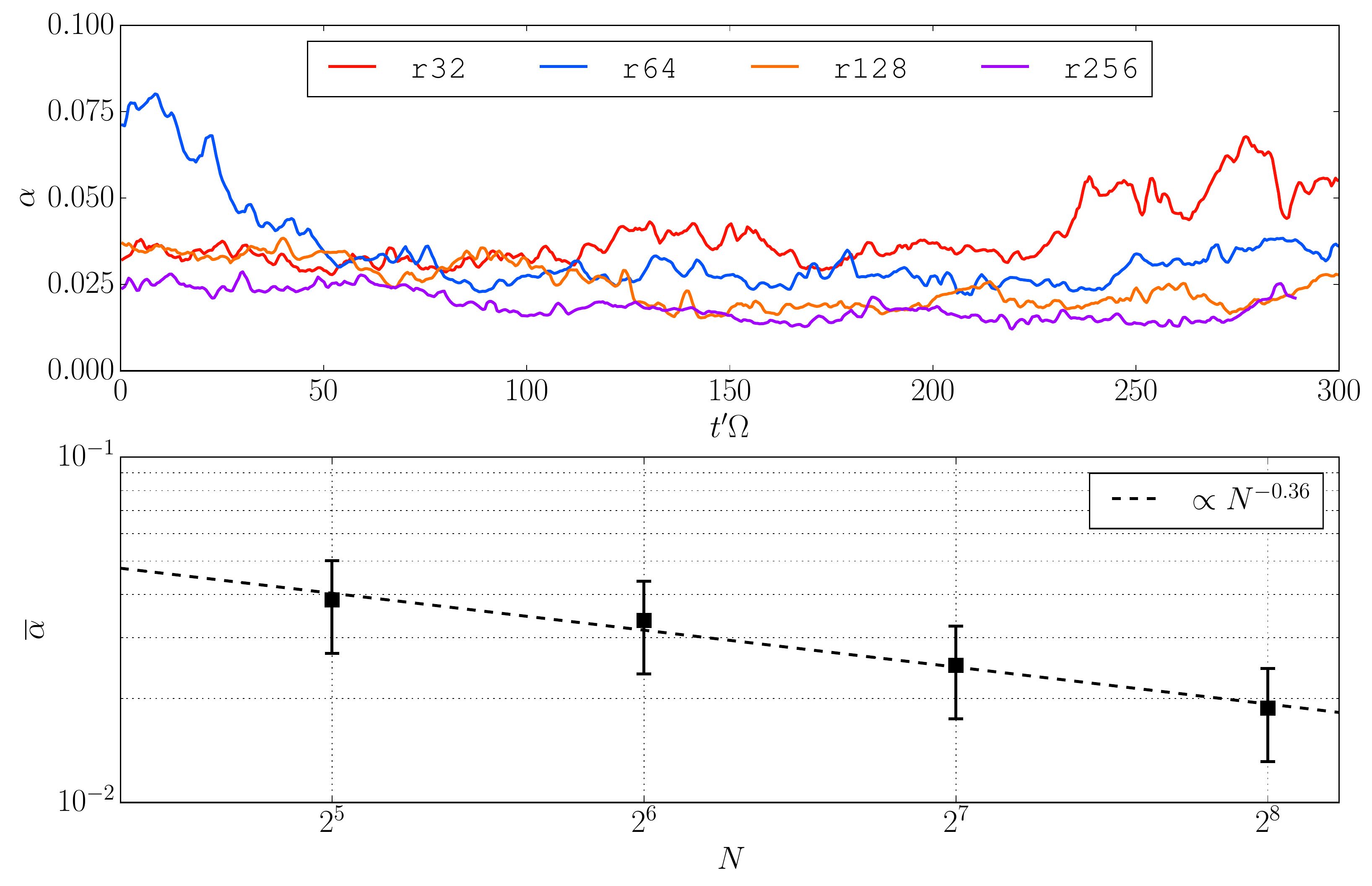}
\caption{$\alpha$ for all runs. The top panel shows evolution over time (after boxcar smoothing of width $\Delta t = 2.5 \Omega^{-1}$ for clarity), while the bottom panel shows the time averages as a function of resolution with a best-fit power law overlaid.}
\label{fig:alpha}
\end{figure}

The run of magnetic field energy density $[E_B](z,t) = \left [{\bf B} \cdot {\bf B} / 2 \right]$ for all runs is shown in Figure \ref{fig:EBspacetime}.  Evidently the ``butterfly'' or dynamo oscillations, which are independent, quasi-periodic enhancements in magnetic energy density on either side of the disk, followed by buoyant rise of magnetic field through $z \sim 2 H$, are present at all resolutions.

\begin{figure}     
\plotone{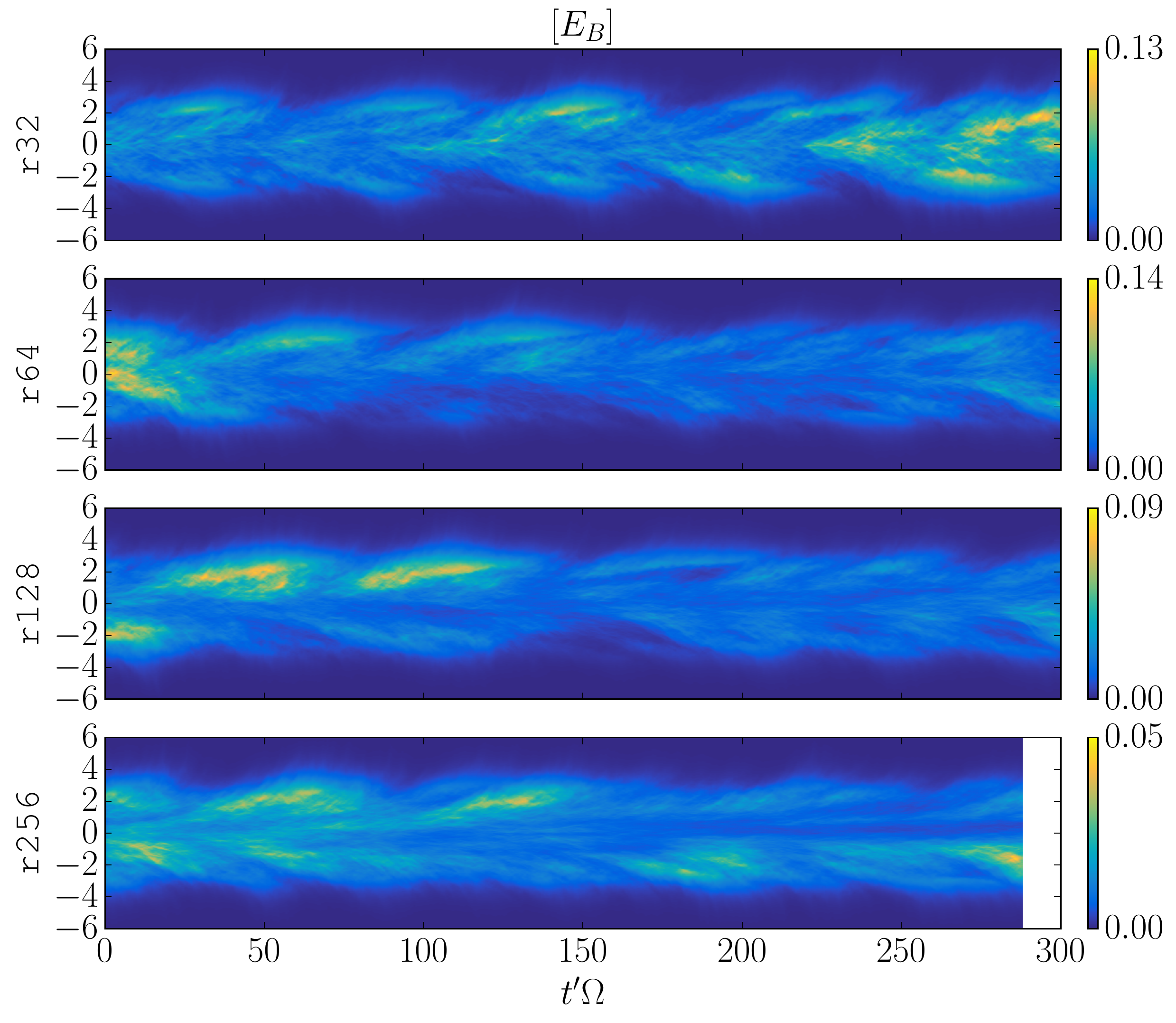}
\caption{Spacetime diagram of $[E_B]$ for all runs. Color scales are specific to each panel. Note persistence of butterfly diagram across all resolutions.}
\label{fig:EBspacetime}
\end{figure}

Does the time-averaged vertical structure of the disk change with resolution? Figure \ref{fig:zprofiles} shows $x$,$y$-averaged quantities for all runs averaged over time. Also shown are fits to $\rho$ and $E_B$ from \cite{guan2011}, who study boxes of lower resolution but greater radial and azimuthal extent than we do here.\footnote{The fit is $\rho = 0.93\rho_0\exp(-z^2/(2 H^2))$ for $|z| < 2.55 H$ and $\rho = 0.036\rho_0 \exp(-(|z|/H - 2.55)/0.44)$ otherwise, and $E_B = 0.012 \rho_0 c_s^2$ for $|z| < 2.55 H$ and $0.012 \rho_0 c_s^2 \exp(-(|z|/H - 2.55)/0.64)$ otherwise.} The density profile is  consistent with an exponential profile (rather than Gaussian) at large $|z|$, with scale height $0.44 H$.  The magnetic energy density is also consistent with an exponential profile at large $|z|$, but with scale height $0.64 H$.  $E_B$ has a feature close to the vertical boundaries, perhaps caused by field lines breaking as they intersect the boundary \citep{millerstone2000}.

The top right panel of Figure \ref{fig:zprofiles} shows the $t$,$x$,$y$-average of total stress. Little variation is seen at large $|z|$, and monotonic decrease of stress with resolution is seen near the midplane. Notice, however, that as resolution increases the structure of averaged stress develops a local minimum around $z = 0$ and a local maximum around $|z| \sim 2 H$.

\begin{figure}      
\plotone{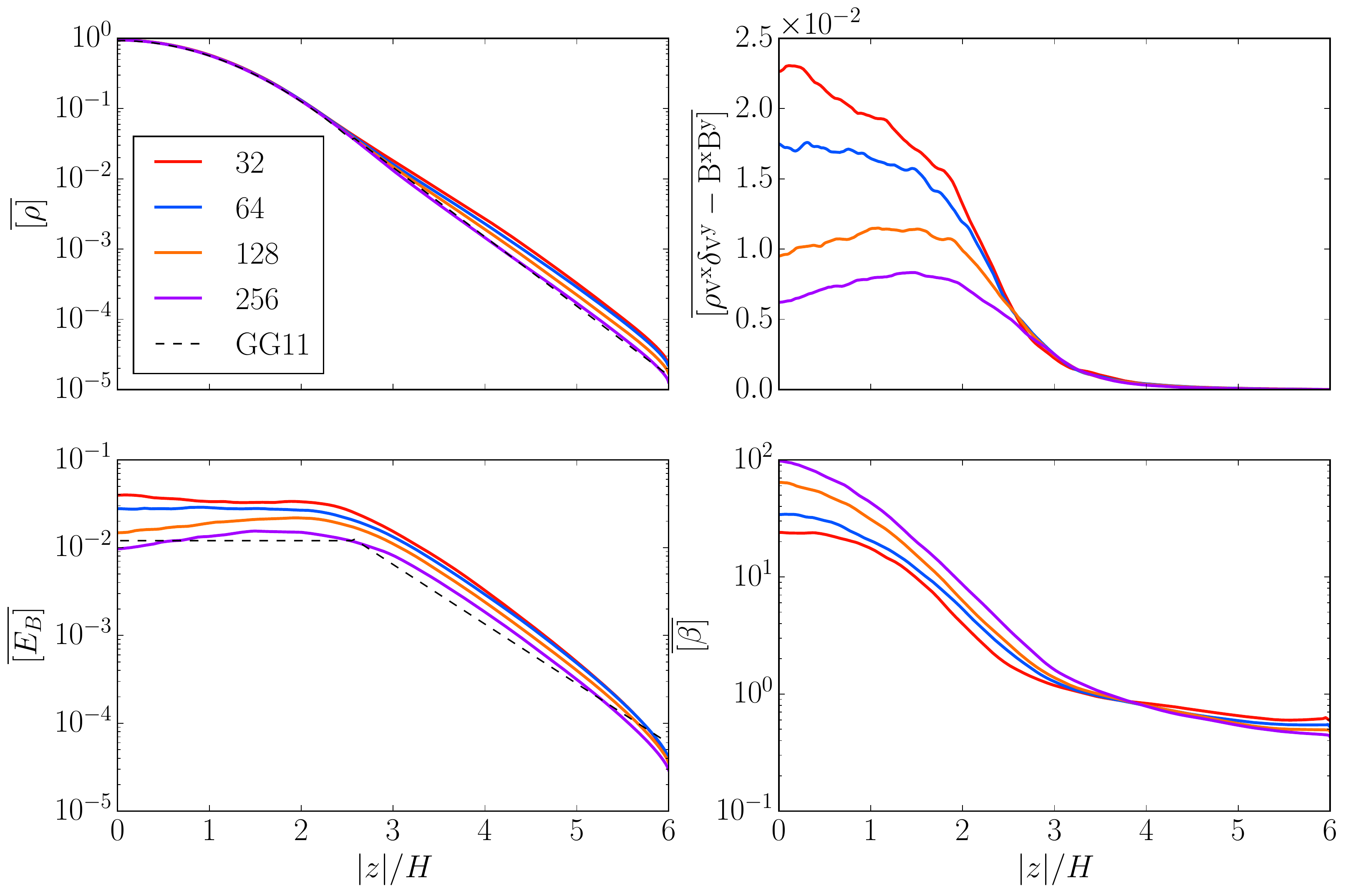}
\caption{Time- and $x$,$y$-averaged quantities as a function of height. Fits to $\rho$ and $E_B$ from \cite{guan2011} are overlaid.}
\label{fig:zprofiles}
\end{figure}

\subsection{Magnetic Field Correlations}
\label{sec:corr}

Earlier work \citep{guan2009} has shown that the magnetic field correlation length (defined below) scales as $N^{-1}$ in zero-net-field, unstratified local ILES models where $\overline{\alpha} \sim N^{-1}$.  How does the characteristic size of structures in MHD disk turbulence change with $N$ for our stratified models?

The dimensionless magnetic field autocorrelation tensor is
\begin{equation}
T^{ij}(\delta x, \delta y, z, t) \equiv \frac{1}{[B^2]}[\delta B^i(x,y,z,t) \delta B^j(x + \delta x, y + \delta y, z, t)].
\end{equation}
The dimensionless scalar magnetic autocorrelation function $\xi_B \equiv Tr(T^{ij})$.  Evidently $\xi_B(\delta x = 0, \delta y = 0) = 1$.  We consider only $\xi_B$; $\xi_v$ and $\xi_\rho$ contain comparatively larger contributions from the compressive disturbances evident in Figure \ref{fig:coolcolorimage} \citep[see also][]{beckwith2011}.

First we average $\xi_B(\delta x, \delta y, z,t)$ over $|z| < 2 H$ and $t$, as did \cite{davis2010}.  The result is shown in Figure \ref{fig:xiB}.  The correlation function is an ellipse swept back by the shear into a trailing spiral structure.  The shape and orientation of the ellipse do not change significantly with resolution, but the scale of the correlation ellipse drops monotonically as resolution is increased.

Next we average $\xi_B(\delta x, \delta y, z,t)$ over time and over bins in $z$ of width $\Delta z = 0.5 H$, then fold around the midplane.  We then evaluate the second moments of $\xi_B(z)$ in the contiguous region around $\delta x = \delta y = 0$ where $\xi_B > 0$.  The eigenvectors of this moment tensor define a major and minor axis with major axis tilted at a small angle $\theta$ to the $y$ axis.  The correlation lengths $\lambda_{\mathrm{minor}}$ and $\lambda_{\mathrm{major}}$ are defined as the distance along each eigenvector at which $\xi_B = 1/e$.  The shape of the correlation departs from an exponential at both small and large scales, although the correlations at large scale are weak and hard to measure accurately (although they must be present, as \cite{guan2011} have shown that butterfly oscillations are coherent over large boxes).  The correlation length is the outer scale of disk turbulence.

Figure \ref{fig:zonesperL_tilt} shows $\lambda_{\mathrm{minor}}(|z|)$, $\lambda_{\mathrm{major}}(|z|)$, and $\theta(|z|)$.  All depend on height.  The tilt rises toward $\approx 19\degree$ for $|z| < 2.5 H$.  It declines out to $4.5 H$ and then rises again toward the boundary (this rise may signal the influence of boundary conditions or the density floor). The major axis correlation length converges toward $\approx 0.5H$ for $|z| > 3H$, but is monotonically decreasing with $N$ at $z = 0$.  The minor axis correlation length is also monotonically decreasing with $N$ at $z = 0$, and rises steadily with a bump at $\approx 3H$ toward the boundaries. 

Figure \ref{fig:convVSz} shows explicitly the resolution sensitivity $d\log\lambda/d\log N$ for the minor and major axis correlation lengths, along with the resolution sensitivity $d\log \overline{[w_{r \phi}]}/d\log N$ of the shear stress, as a function of $|z|$. Both correlation lengths are sensitive to resolution at the midplane, and far less sensitive (perhaps converged) at higher altitude.  At the midplane, both correlation lengths scale as $N^{-1/2}$. $\overline{[w_{r \phi}]}$ exhibits a similar trend, especially for $|z| \lesssim 3H$.

Does this mean the outer scale of turbulence is unresolved, even at our highest resolution?  Figure \ref{fig:zonesperL_tilt} also shows $\lambda_{\mathrm{minor}}(|z|)$, $\lambda_{\mathrm{major}}(|z|)$ in units of $\Delta x$ in the right panels.  Above $|z| = 3H$ even the minor axis is very well resolved, with in excess of $30$ zones per correlation length.  At the midplane $\lambda_{\mathrm{minor}}(N = 32)/\Delta x \approx 3$ and $\lambda_{\mathrm{minor}}(N = 256)/\Delta x \approx 10$.  This differs from the nonconvergence observed in unstratified, zero-net-field ILES models, where $\lambda/\Delta x$ are independent of $N$; here, the outer scale is better resolved as resolution increases.\footnote{The ratio of correlation length to resolution $\lambda/\Delta x$ is related to, but not exactly the same as, the quality factor $Q \equiv \lambda_{MRI}/\Delta x$, where $\lambda_{MRI} \propto v_A/\Omega$ is a characteristic wavelength for the MRI \citep{sano2004,noble2010,hawley2011}.  The ratio of the two ratios is $\propto M \equiv v_A/ \Omega \lambda$, which is the Alfv\'en Mach number of MRI-driven turbulence at the correlation length. \cite{walker2016} demonstrated that in their unstratified models $M$ is approximately constant in MRI driven turbulence.  In our simulations $M$ varies by a factor $\sim 2$ inside the disk.}

\begin{figure}      
\plotone{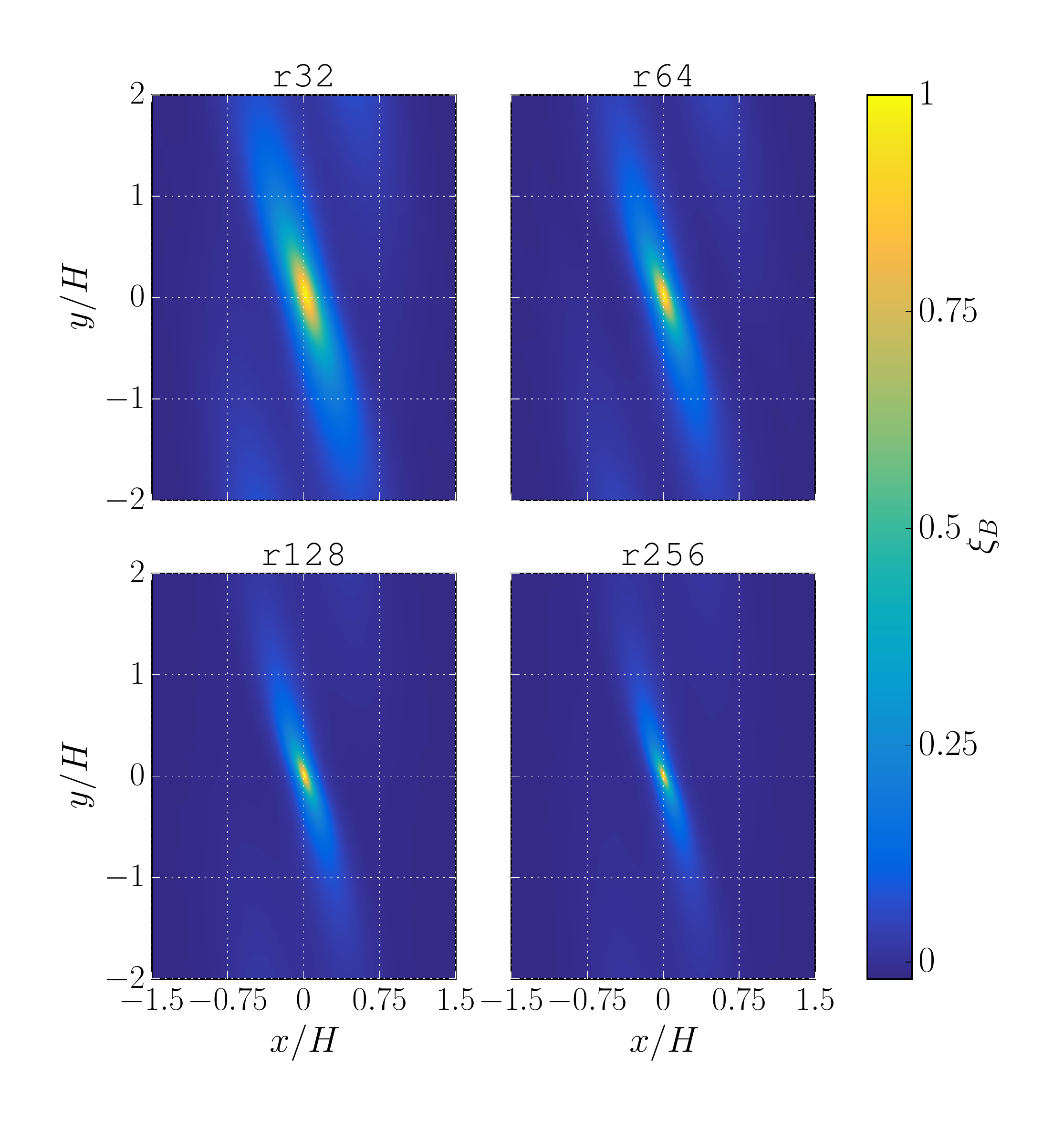}
\caption{$\xi_B$ averaged over time and the region $|z| < 2H$ for all runs.}
\label{fig:xiB}
\end{figure}

\begin{figure}
\plotone{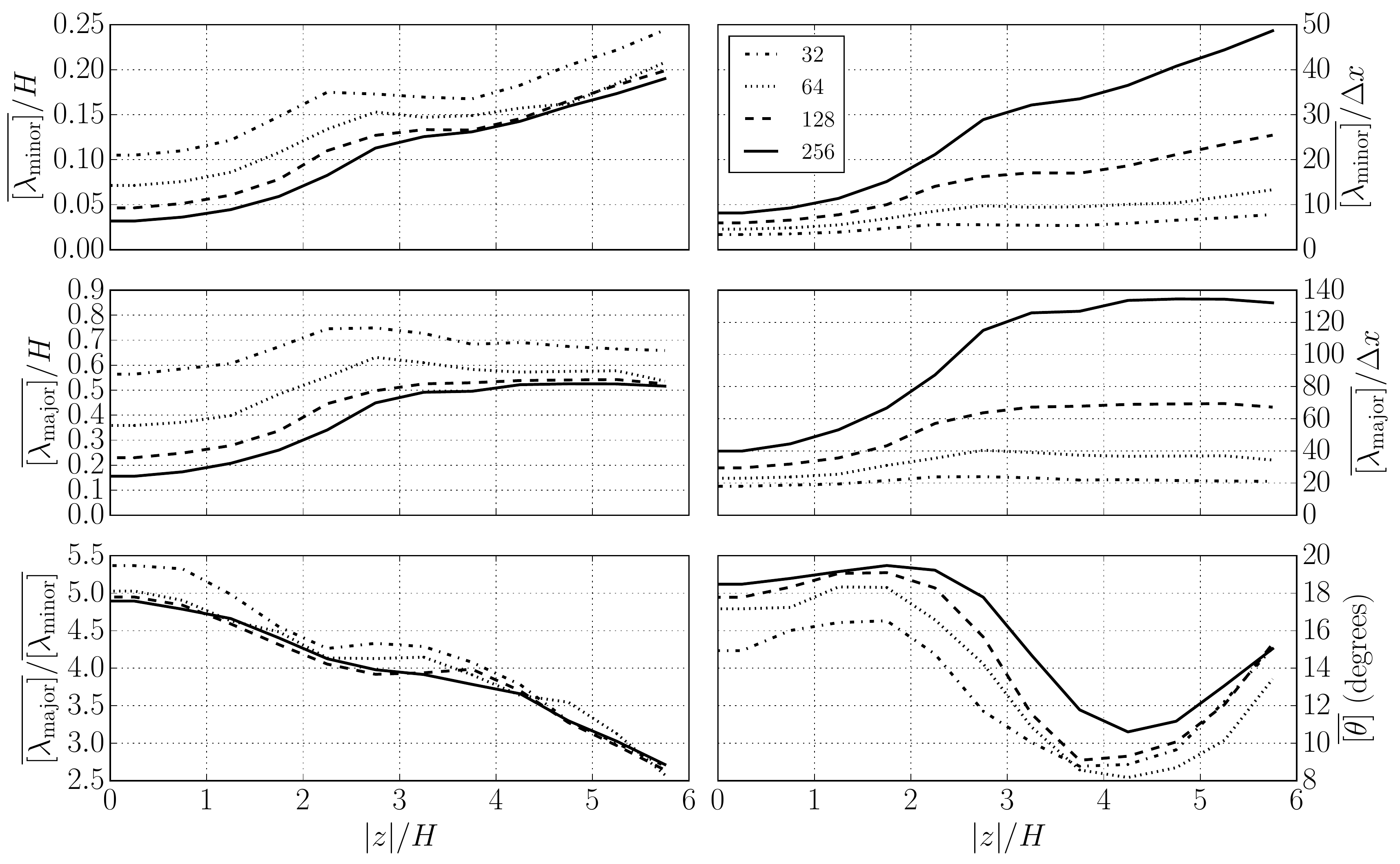}
\caption{Minor axis correlation length, major axis correlation length, and tilt as a function of $|z|$ for each resolution.  The correlation lengths are given in units of scale heights (left panels) and cell size $\Delta x$ (right panels).}
\label{fig:zonesperL_tilt}
\end{figure}

\begin{figure}      
\plotone{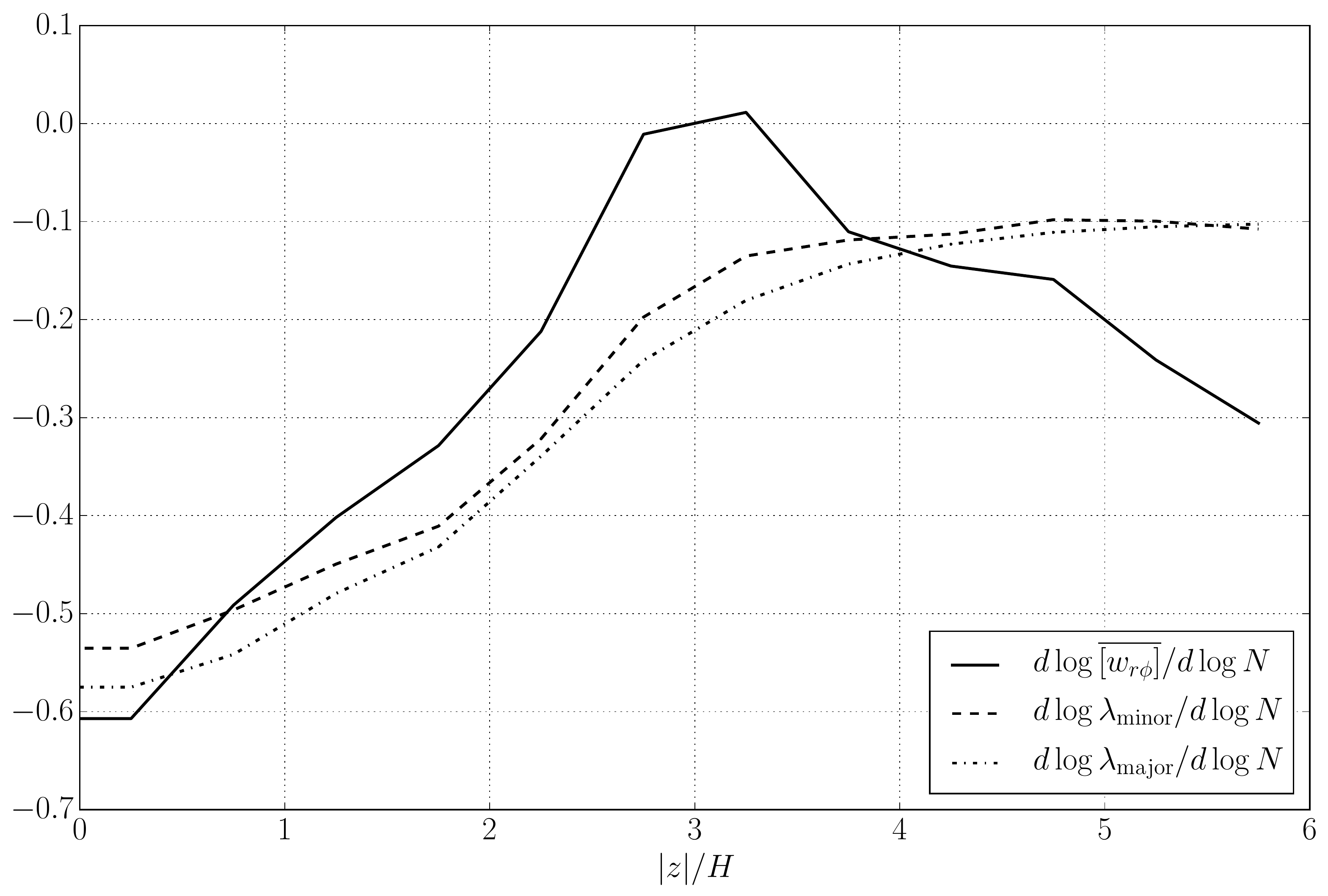}
\caption{Resolution dependence of shear stress $d\log \overline{[w_{r \phi}]}/d\log N(z)$ and correlation lengths $d\log \lambda/d\log N(z)$ for $\lambda$ the minor axis (middle) and major axis (bottom) magnetic field correlation lengths. Both $\lambda$ are more strongly dependent on resolution at the midplane than at $|z| > 3 H$.}
\label{fig:convVSz}
\end{figure}

\subsection{Evolution of Net Magnetic Flux}

\begin{deluxetable}{cccccc}
% No blank lines inside deluxetable
%\tabletypesize{\footnotesize} % 11 pt, \footnotesize for 10pt, nothing for 12pt
\tablewidth{0pc}
\tablehead{
\colhead{Label} &
\colhead{$\overline{\langle B^x \rangle}_{\mathrm{RMS}}$} &
\colhead{$\sigma_{{\langle B^x \rangle}}$} &
\colhead{$\overline{\langle B^y \rangle}_{\mathrm{RMS}}$} &
\colhead{$\sigma_{{\langle B^y \rangle}}$} &
\colhead{$\overline{\alpha}_{\mathrm{NF}}$}
}
\tablecaption{RMS and standard deviation of net magnetic fluxes present for each run. Vertical magnetic flux is zero, conserved to machine precision.\label{tab:netflux}}
\startdata
    {\tt r32}  & $7.0\times10^{-4}$  & $4.2\times10^{-4}$ & $3.6\times10^{-2}$ & $1.5\times10^{-2}$ & $1.1\times10^{-2}$ \\
    {\tt r64}  & $4.0\times10^{-4}$  & $3.8\times10^{-4}$ & $1.8\times10^{-2}$ & $1.6\times10^{-2}$ & $5.6\times10^{-3}$ \\
    {\tt r128} & $4.3\times10^{-4}$ & $3.2\times10^{-4}$ & $2.0\times10^{-2}$ & $1.4\times10^{-2}$ & $6.2\times10^{-3}$ \\
    {\tt r256} & $3.7\times10^{-4}$ & $3.5\times10^{-4}$ & $1.1\times10^{-2}$ & $1.0\times10^{-2}$ & $3.4\times10^{-3}$ \\
\enddata
\end{deluxetable}

Our choice of boundary conditions permit evolution of $\langle B^x \rangle$ and $\langle B^y \rangle$. How important is the mean field in driving the evolution?

The RMS and standard deviation of $\< B^y \>$ and $\< B^x \>$ are given in Table \ref{tab:netflux}.  Evidently $\< B^y \> \gg \< B^x \>$.  We can estimate the effect of the mean field on $\overline{\alpha}$ using the saturation predictor of \cite{HGB} for an unstratified shearing box with a net toroidal field\footnote{We emphasize that this predictor is for {\it unstratified} models; how well it recovers the behavior of stratified models is uncertain. We also use the mean field through the box as input; locally, the net field may vary.}: $B^2\sim 4 \pi \sqrt{16/15} (0.012) \rho_0 L_y v_A \Omega$, where $v_A$ is the \alfven{} speed associated with the rms net toroidal field.  Then using $\overline{\alpha}_{NF} \approx \int dz (1/(2\beta))/\int dz P = 0.25 B^2 (5 H) / \sqrt{2 \pi}$ (where $5H$ comes from assuming $B^2 = \mathrm{const}$ for $|z| < 2.5H$, $B^2=0$ else), we find a predicted $\overline{\alpha}$ associated with the mean azimuthal field that is, for all models, at least a factor of $3$ smaller than the measured $\overline{\alpha}$ (and nearly an order of magnitude for {\tt r256}).  This suggests that the boundary conditions are not controlling the saturation.

The mean field sensed locally by the turbulence may still control $\alpha$ locally.  To illustrate this point, Figure \ref{fig:netabsoluteflux} shows a sample estimate of a local mean field: the azimuthal field averaged over sheets at constant $z$.  This fluctuates in sign, so to avoid cancellation we take the time average of the absolute value of this mean field.  The resulting mean field is an order of magnitude larger than $\<B_y\>$, which the unstratified box saturation predictor suggests would produce an $\overline{\alpha}$ comparable to what is measured.  In sum: a localized mean field may play an important role in controlling the outcome, but the mean field over the entire computational domain does not.

\begin{figure}
\plotone{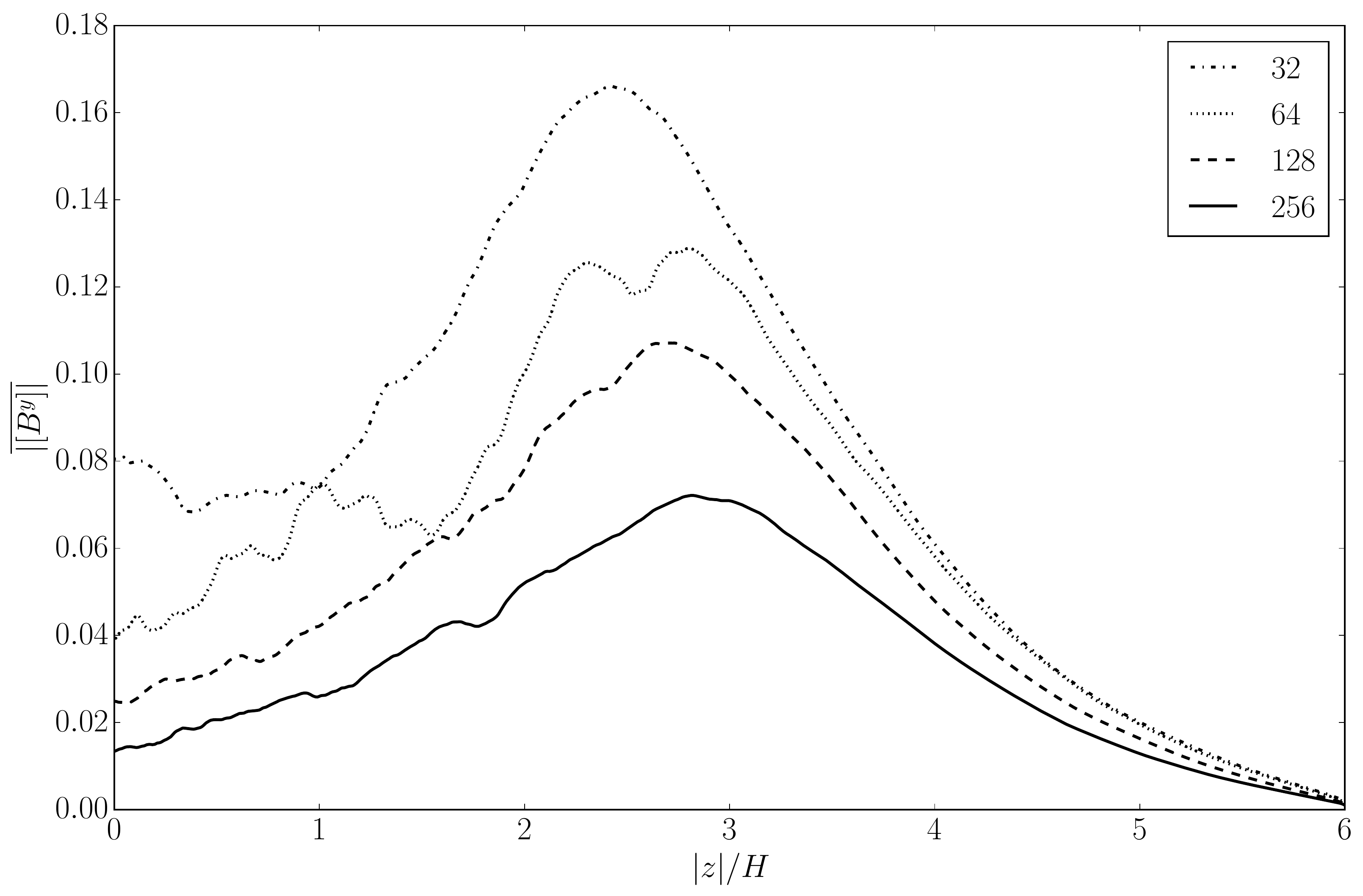}
\caption{Time-averaged absolute value of the $x$,$y$-averaged magnetic flux.}
\label{fig:netabsoluteflux}
\end{figure}

\section{Discussion}
\label{sec:discussion}

Our simulations have yielded several unexpected dependences on resolution: (1) $\overline{\alpha} \sim N^{-1/3}$ (2) $\lambda_{\mathrm{minor}} \propto \lambda_{\mathrm{major}} \sim N^{-1/2}$ in the midplane of the disk, and nearly $\sim N^0$ in the corona (3) The total stress, scaling similarly to $\lambda$ with $N$, develops a local maximum at $|z| \sim 2H$ as resolution increases.

Surprisingly, we do not see convergence of the time-averaged, vertically integrated shear stress $\overline{\alpha}$ for resolution up to $N = 256$ zones per scale height in stratified, isothermal, local ILES models. This is broadly consistent with \cite{bodo2014} and in tension with the results of \cite{davis2010}. Both \cite{davis2010} and \cite{bodo2014} find a plateau in $\overline{\alpha}$ between $\approx 45$ and $\approx 90$ zones per $H$. We do not find evidence for this behavior, but the plateau could be hidden in our measurement errors due to finite run time and finite computational volume.

Are our results consistent with earlier work?  To compare, we need to convert to common units and a common measurement of stress, for which we will use $\alpha$ as defined in eq.(\ref{eqn:alphadef}).

\cite{davis2010} report volume-and-time averaged stresses in units of the midplane pressure.  This is equivalent to volume-averaged stress in our units.  Notice that \cite{davis2010} define $H = \sqrt{2} c_s/\Omega$.  Then for $N \approx (23,$ $45,$ $91)$ their volume averaged stress (see their Table 1) is $(0.0149,$ $0.0093,$ $0.0092)$.  Converting to vertically integrated stress (multiply by $4\sqrt{2}$) and dividing by the vertically integrated gas pressure ($\sqrt{2\pi}$ in our units), we find (using our definition of $\alpha$) $\overline{\alpha} = (0.034,$ $0.021,$ $0.021)$.

\cite{bodo2014} also define $H = \sqrt{2} c_s/\Omega$, and set $c_s = 1/\sqrt{2}$, $\rho_0 = 1,$ and $\Omega = 1$, so their unit of stress is a factor of $2$ larger than ours.  They consider models with $N \approx (23,$ $45,$ $91,$ $141)$.  Since they do not report time-averaged stresses, we will estimate these from their Figure 2.  We estimate that the volume integrated maxwell stress in their units is $\simeq (0.022,$ $0.017,$ $0.017,$ $0.01)$.  We convert this vertically integrated stress to our units (multiply by $2 \sqrt{2}$; the factor of $2$ is for the stress unit and the factor of $\sqrt{2}$ is for the length unit), multiply by $1.25$ to incorporate an assumed $25\%$ Reynolds stress contribution, then divide by the vertically integrated pressure ($\sqrt{2\pi}$) in our units to find $\overline{\alpha} \simeq (0.031,$ $0.024,$ $0.024,$ $0.014)$.  

To facilitate comparison, at a resolution of $N = (32,$ $64,$ $128,$ $256)$ we find $\overline{\alpha} = (0.039,$ $0.034,$ $0.025,$ $0.019)$.  These results are shown in Figure \ref{fig:alphacomp}.  The overall offset of the Davis et al. and Bodo et al. series from ours is significant, but may be explained in part by the larger vertical extent of our models.  The algorithms used also differ, possibly yielding different effective resolutions, and of course the vertical boundary conditions also differ.  Nevertheless, it is reassuring that all simulations lead to values of $\overline{\alpha}$ that are within $1\sigma$ of our results.  Indeed, least squares power-law fits to the Davis et al. and Bodo et al. series yield slopes ($-0.35$ and $-0.37$, respectively) consistent with ours ($-0.36$) and the relationship $\overline{\alpha} \sim N^{-1/3}$.

\begin{figure}
\plotone{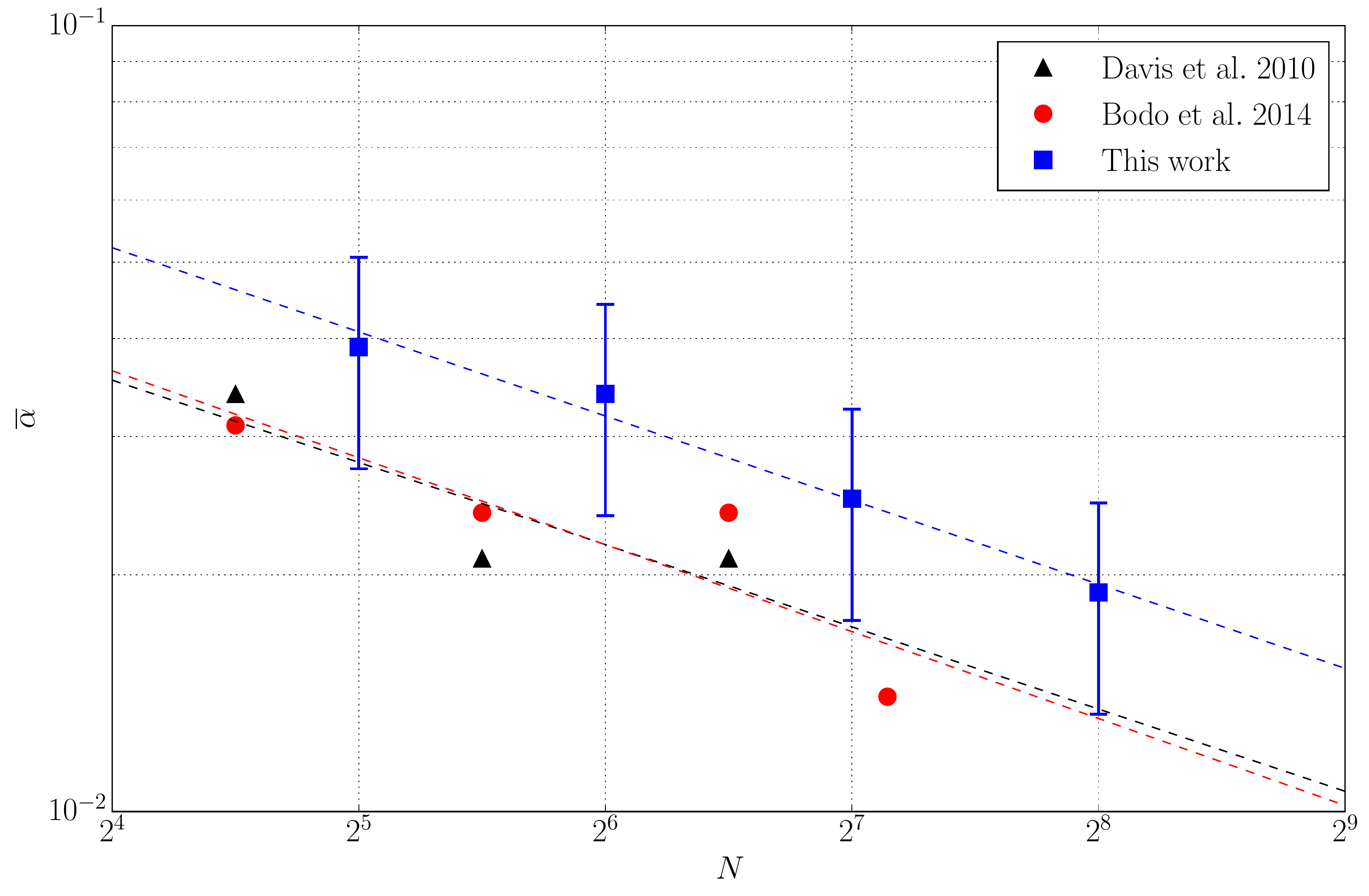}
\caption{Time-averaged dimensionless shear stress $\overline{\alpha}$ for \citealt{davis2010}, \citealt{bodo2014}, and this work. Results are broadly consistent, and all show approximately the same scaling of stress with resolution. Fits to each dataset are shown as dashed lines.}
\label{fig:alphacomp}
\end{figure}

The correlation function in the $x-y$ plane is approximately ellipsoidal and characterized by the major axis length, minor axis length, and the ``tilt angle'' between the major axis and the $y$ axis.  The tilt angle $\theta_{\mathrm{tilt}}(N = 256) \sim 19\degree$ is consistent with \cite{davis2010} who find $\theta_{\mathrm{tilt}}(N \approx 91) \sim 18\degree$. The increase in $\theta_{\mathrm{tilt}}$ with resolution was also reported by \cite{guan2009}. Although \cite{davis2010} do not quote a value for $\lambda_{\mathrm{minor}}$, visual inspection of their $\xi_B(N\approx64)$ slice yields a value comparable to what we find at similar resolution.

The sensitivity of stress to $N$ depends on height (see Figure \ref{fig:convVSz}).  The midplane shear stress decreases with $N$ at a rate that is inconsistent with convergence, but the stress at $|z| \gtrsim 2 H$ is much less sensitive to $N$ and convergence is not excluded by our limited time- and volume-sampled data. One consequence of this is that a local minimum develops in the total stress at $z = 0$ and a local maximum develops at $|z| \simeq 2 H$.  A qualitatively similar local maximum in the stress is observed in stratified shearing boxes with self-consistent thermodynamics, at least when they are radiation pressure-dominated (\citealt{hirose2009}, \citealt{jiang2016}). This effect appears to be due to a convective process which also significantly enhances $\alpha$ in these models \citep[e.g.][]{hirose2014}.

Are our simulations run long enough?  From a long-duration, low-resolution simulation we measured a correlation time of $\approx 60 \Omega^{-1}$ (this is slightly shorter than the $90\Omega^{-1}$ correlation time seen in the $N = 90$ run of \cite{davis2010} \footnote{We thank S. Davis for kindly providing us with the data.}), and our assessment of the error bars on $\bar{\alpha}$ relies on this measurement.  Stratified shearing box models frequently give an impression of order-unity enhancements in $\alpha$ (``bursts'') separated by long intervals, and rare bursts could change the correlation time.  Our data are not sufficient to assess whether this impression is statistically well grounded or not.  If it is, then the bursts might correspond to long-timescale power in the power spectrum of a Gaussian process that is undetectable in a short simulation, or non-Gaussianity associated with the flares.  There is, however, no evidence for non-Gaussianity in our data; the probability distribution for $\log{\alpha}$, for example, is consistent with Gaussian.  There is also no evidence to changes in the variance of $\log(\alpha)$ with $N$; the relative variance, shown in Table 3, shows no systematic trend.

Why no convergence? The cause may lie either with our numerical realization of the stratified isothermal zero-net-flux ILES shearing box model (A), or with assumptions made by the model itself (B). We have assembled an incomplete list of possible explanations:

(A1) The nonconvergence is physical and $\alpha_{MRI} \rightarrow 0$ in isothermal astrophysical disks with vanishing mean field.  Although we cannot rule this out, it seems inconsistent with the result of \cite{fromang2010} for an unstratified model with explicit scalar viscosity and resistivity that converges to nonzero $\alpha$, albeit only for $P_m = 4$. 

(A2) The apparent nonconvergence is a consequence of a combination of statistical errors associated with a finite sampling time and an initial transient that results from using resolution $N/2$ data to initialize resolution $N$ models.  Our analysis (see Appendix) suggests, however, that even though $\overline{\alpha}$ has a long correlation time this is improbable.  

(A3) The nonconvergence is an artifact of the limited size of the model.  Fluctuations in $\alpha(t)$ will depend on the volume of the simulation.  Naively, they would scale as $1$ over the square root of the number of correlation volumes.  But there is coupling between correlation volumes via large-scale magnetic fields and this is connected to the butterfly oscillations.  Furthermore, it is already known that in unstratified, local simulations the imposition of a mean field causes an ILES model to converge.  Ultimately, it must be that turbulence is locally unable to distinguish between uniform fields and magnetic fields that have structure on a sufficiently large scale.  Perhaps our models are simply too small to see this sufficiently large scale, and so they are analogous to the zero mean field unstratified models that do not converge.  

The interaction between small and large scale fields has been  explored by \cite{sorathia2012} who measured the net magnetic flux in local regions of global unstratified ILES models. They found distributions of $\langle B^x \rangle$, $\langle B^y \rangle$, and $\langle B^z \rangle$ inconsistent with zero, with the linear MRI growth associated with these mean fields typically being well resolved in their simulations.

(A4) The model will converge at higher $N$, and we are simply not in the high resolution limit yet.  The magnetic field correlation length is $\sim 10 \Delta x$ in our highest resolution models, so there is only a dynamic range of $\simeq 2$ between the outer scale $\lambda$ and the dissipation scale.  

(A5) The model will {\it not} converge, with $\alpha \sim \lambda \sim N^{-1}$, in the complete absence of a mean magnetic field. Although we cannot account for the $\overline{\alpha}$ we measure from the net flux through our computational domain, our estimate is based on a fit to results from unstratified models. Mean fields in stratified boxes may behave differently. They may, for example, be playing a stronger role than we estimate in the magnetically-dominated corona, contributing to our near-convergence of $\lambda$ above $|z| \sim 3H$. Nonetheless, \cite{davis2010} {\it do} maintain zero net flux in a stratified model, and their results are inconsistent with $\alpha \sim N^{-1}$.

(B1) The nonconvergence is an artifact of our use of an ILES model. In models in which the numerical resolution and Reynolds numbers are increased together, there is numerical evidence that both unstratified and stratified models converge (\citealt{fromang2010}, \citealt{simon2011}). It would be interesting to know whether this extends to larger $N$ and the large $Re_M$, large $Re$ limit relevant to astrophysical disks. There is also numerical evidence that computational models of the solar dynamo depend strongly on the dissipation model \citep[see, e.g.][for a review]{Charbonneau2014}.

(B2) The nonconvergence is an artifact of the absence of consistent vertical energy transport by radiation and convection.  It is now known that convective disks in models with consistent treatment of energy transport exhibit enhanced $\alpha$ (\citealt{hirose2014}), which may enhance the amplitude of dwarf nova outbursts. The convective process may aid convergence (\citealt{bodo2015}).  It is not yet clear how well converged the energetically consistent models are; current model have $N \sim O(64)$. 

(B3) The nonconvergence is an artifact of the symmetry of the local model.  The local model is invariant under translations in the plane of the disk, and invariant under rotations by $\pi$ around the $z$ axis.  The incorporation of higher order terms in $H/r$ would break these symmetries and might qualitatively change the outcome.  There is limited numerical evidence for convergence in unstratified global models \citep{sorathia2012}, although with a tendency for $\overline{\alpha}$ (and hence $\dot{M}$ and $\beta^{-1}$) to increase with resolution \citep[see also][]{shiokawa2012, hawley2011, hawley3D}.

We are unsure which (if any) of these explanations is correct, but all except (A1) are amenable to future numerical investigation.

What are the implications of nonconvergence?  It is difficult to say without testing the hypotheses above with new numerical simulations.  For example, if A4 is correct (insufficient resolution) then current lower resolution models may yield $\bar{\alpha}$ to within a factor of two.  On the other hand, if A5 is correct ($\alpha$ is zero without a mean field) then the result would have profound implications for our understanding of disk structure and evolution, which would presumably be controlled by the generation and transport of large-scale magnetic field.  No matter what the explanation for the nonconvergence seen here, future disk simulations need to be tested carefully for convergence.

\section{Conclusion}

The isothermal stratified zero-net-flux shearing box is a minimal model with zero physical parameters for the turbulent saturation of the magnetorotational instability and is thus central to accretion disk theory.  We have attempted to sort out apparently conflicting reports of convergence in the literature using the {\tt ramses-gpu} code on \bluewaters{} to probe convergence at an unprecedented resolution of $N = 256$ zones per scale height.

Our results imply that existing local and perhaps global zero-mean-field ILES models of disks are, at best, underresolved.  We have found that $\overline{\alpha} \sim N^{-1/3}$.  This is not convergent, but it differs from the sharp nonconvergence identified by \cite{fromang2007} in unstratified ILES models, with $\overline{\alpha} \sim N^{-1}$.

We have also compared our results to earlier work by \cite{davis2010} and \cite{bodo2014}.  These earlier calculations are consistent with our to within the error bars, and all show a similar trend with resolution. Like \cite{bodo2014} and unlike \cite{davis2010}, our models do not conserve net toroidal magnetic flux. Although first estimates suggest the net flux present in our model is not controlling our results, this remains an uncertainty in performing comparisons. Box size effects may also confound comparisons.

We have reviewed possible physical and numerical causes of this nonconvergence.  All of these are amenable to further numerical investigation when sufficient computational resources are available. One implication is clear, however: simulations of MHD turbulence in disks need to be tested carefully for convergence, and the attendant uncertainties need to be allowed for when weighing the results.

\acknowledgments
It is a pleasure to thank G. Bodo, M. Chandra, S. Davis, J. Dolence, V. Paschalidis,  J. Simon, and J. Stone for useful discussions. The numerical calculations presented here were performed on the Blue Waters supercomputer at NCSA.  BRR was supported by an Illinois Distinguished Fellowship and by NSF grant AST-1333612. Work at Los Alamos National Laboratory was done under the auspices of the National Nuclear Security Administration of the US Department of Energy. CFG's work was also supported in part by a Romano Professorial Scholar appointment, a Simons Fellowship in Theoretical Physics, and a Visiting Fellowship at All Souls College, Oxford.

\clearpage

\appendix
\section{Measurement Error Estimates with a Gaussian Process Model}

Shearing box simulations estimate the true, long-term average $\alpha_0$ from $\overline{\alpha}$ measured over a finite time $\Delta t$.  How long is long enough?  

Note that in this section, $\< \>$ denotes an expectation value for consistency with previous literature on Gaussian random fields, rather than the volume average of Equation \ref{eqn:volavg}. Suppose $\alpha$ has a correlation time $\tau_c$ and variance $\sigma^2$.  Then our intuition is that $\<(\overline{\alpha} - \alpha_0)^2\>$should be proportional to $\sigma^2 \tau_c/\Delta t$, i.e. the rms error averaged over many realizations of $\alpha$ should scale as one over the square root of the number of correlation times.  But with what coefficient?

We can estimate $\<(\overline{\alpha} - \alpha_0)^2\>$ for a Gaussian process with known power spectrum over some long but finite time $T$.  That is,  
\begin{equation}
\alpha(t) = \alpha_0 + \sum_j a_j \cos(\omega_j t + \phi_j); \qquad \omega_j = \frac{2\pi j}{T}
\end{equation}
The sum is taken only over $\omega_j > 0$, $\phi_j$ is uniformly distributed in $[0,2\pi)$ (random phase) and $a_j$ is Gaussian distributed:
\begin{equation}
P(a_j) d a_j  = \exp\left(-\frac{a_j^2}{2 \<a_j^2\>}\right) \frac{d a_j}{(2 \pi \<a_j^2\>)^{1/2}}
\end{equation}
The power spectrum $P_\omega \equiv  T \<a_j^2\>$.  In the limit that $T$ is large the modes are closely spaced and $\sum_j \rightarrow \int d\omega T/(2 \pi)$.  The expected variance in $\alpha$ over the interval $T$ is
\begin{equation}
\<\sigma^2\> = \< (\alpha - \alpha_0)^2 \> = \frac{1}{2} \int_0^\infty \frac{d\omega}{2\pi} P_\omega.
\end{equation}
where the factor of $2$ comes from phase averaging.
$P_\omega$ is independent of $T$ if $\sigma^2$ is fixed. 

The autocorrelation function is
\begin{equation}
\xi (\tau) \equiv \frac{1}{T} \int_0^T \, dt \, \alpha(t)\alpha(t + \tau) - \alpha_0^2 = \frac{1}{2} \int_0^\infty \, \frac{d\omega}{2\pi} \, P_\omega \cos\omega\tau.
\end{equation}
The error in estimating $\alpha_0$ from a finite interval $\Delta t$ is
\begin{equation}
m = \frac{1}{\Delta t} \int_0^{\Delta t} \, dt \, (\alpha - \alpha_0)
\end{equation}
Expanding and integrating,
\begin{equation}
m = \frac{1}{\Delta t} \int_0^{\Delta t} dt \sum_j a_j \cos(\omega_j t + \phi_j) =
\sum_j a_j \frac{1}{\omega_j \Delta t} \left\{\sin(\omega_j \Delta t + \phi_j) -
\sin\phi_j\right\}
\end{equation}
Then
\begin{equation}
\< m^2 \> = \sum_j \< a_j^2 \> \frac{1}{\omega_j^2 \Delta t^2} \left(1 - \cos(\omega_j \Delta t)\right)
\rightarrow \int \frac{d\omega}{2\pi} \frac{P_\omega}{\omega^2 \Delta t^2}
\left(1 - \cos(\omega \Delta t)\right).
\end{equation}
To go further we need to know the power spectrum.

We consider model power spectra that decorrelate on long timescales, so that $P_\omega \propto \omega^0$ for $\omega$ small, and scale as a power law at high frequency.  A suitable model is
\begin{equation}
P_\omega \propto (1 + (\omega/\omega_0)^2)^{-p/2}.
\end{equation}
Evidently if the process is stationary then $p > 1$.  The power spectrum can be normalized by $\sigma^2$:
\begin{equation}
P_\omega = \frac{\Gamma(p/2) \, 8 \sqrt{\pi} }{\Gamma((p-1)/2)}
\frac{\sigma^2}{\omega_0} \frac{1}{(1 + \omega^2/\omega_0^2)^{p/2}}.
\end{equation}
Then
\begin{equation}
\xi(\tau) = \frac{ 2^{3/2 - p/2} }{\Gamma((p-1)/2)} \,\, \sigma^2 \,\, (\omega_0 |\tau|)^{(p-1)/2}
K_{(p-1)/2}(\omega_0 |\tau|)
\end{equation}
where $K_n$ is a modified Bessel function of the second kind.  It is easy to show that $\xi(0) \rightarrow \sigma^2$.  

We estimate $\sigma^2$ from data taken over an interval $\Delta t$.  This estimate is biased because it does not include contributions to the variance from low frequency components.  The expected value of $\sigma^2$ sampled over time $\Delta t$ is
\begin{equation}
\< (\alpha - \overline{\alpha})^2 \> = 
\int_0^\infty \frac{d\omega}{2\pi} P_\omega 
\left(\frac{1}{2} + \frac{\cos(\omega\Delta t) - 1}{\omega^2\Delta t^2}\right).
\end{equation} 
If $p$ and $\omega_0$ are known then this expression can be used to produce an debiased estimate of $\sigma^2$. 

An auxiliary $N = 32$ run with $\Delta t = 2000\Omega^{-1}$ has a power spectrum consistent with $p \sim 2$.  For this special case, $\xi(\tau) = \sigma^2 e^{-\omega_0 |\tau|}$, $\tau_c = \omega_0^{-1}$, and 
\begin{equation}
\< m^2 \> = \frac{\sigma^2}{\omega_0 \Delta t} \, 2 \left(1 - \frac{1 -
e^{-\omega_0 \Delta t}}{\omega_0 \Delta t}\right).
\label{eq:Emsq}
\end{equation}
Consistent with expectations, this scales as $\sigma^2 \tau_c/\Delta t$.  Also,
\begin{equation}
\< (\alpha - \overline{\alpha})^2 \> = 
\sigma^2 \left(
1 
- \frac{2}{\omega_0\Delta t} 
+ \frac{2 (1 - e^{-\omega_0\Delta t})}{\omega_0^2 \Delta t^2} 
\right).
\label{eq:sigbias}
\end{equation}
The auxiliary $N = 32$ run has $\tau_c\Omega \simeq 60$, so $\omega_0 = 60^{-1}\Omega$, which we will assume is independent of $N$.  Runs in Table 2 with $\Delta t = 300\Omega^{-1}$ therefore have $\omega_0 \Delta t \simeq 5$.  Runs in Table 2 also have $\sigma_\alpha/\overline{\alpha} \simeq 0.25$.  Then (\ref{eq:sigbias}) implies the debiased $\sigma_\alpha/\alpha \simeq 0.30$.  Combined with (\ref{eq:Emsq}), we find $\<m^2\>^{1/2} = 0.17 \alpha$.  This implies that the one-sigma error is small compared to the total change in $\overline{\alpha}$ over a factor of $8$ in $N$.  

\clearpage

\newpage

% Test for unused citations
\nocite{*}

\bibliographystyle{apj}
\bibliography{local}

\end{document}